\newcommand{\CLASSINPUTinnersidemargin}{25mm}
\newcommand{\CLASSINPUToutersidemargin}{25mm}
\newcommand{\CLASSINPUTtoptextmargin}{35mm}
\newcommand{\CLASSINPUTbottomtextmargin}{20mm}
\RequirePackage[ngerman=ngerman-x-latest]{hyphsubst}
\documentclass[journal, a4paper, twoside]{IEEEtran_MPC}
\PassOptionsToPackage{cmyk}{xcolor}
\usepackage{pst-all}
\usepackage{fix-cm}
\definecolor{MPC_gelb}{cmyk}{0.06,0,0.96,0}
\definecolor{MPC_blau}{cmyk}{1.00,0.3,0,0}
\definecolor{MPC_rot}{cmyk}{0.12,1.0,1.0,0.04}
\usepackage[utf8]{inputenc}
\usepackage[T1]{fontenc}
\usepackage{eurosym}
\usepackage[ngerman]{babel}
\usepackage{blindtext}

\ifCLASSINFOpdf
  \usepackage[pdftex]{graphicx}
  \graphicspath{{.}}
  \DeclareGraphicsExtensions{.jpeg,.jpg,.png,.eps}
\fi

\usepackage{amsmath}
\usepackage{amssymb}
\usepackage{bm}
\usepackage{csquotes}
\usepackage{pifont}
\usepackage{pdfpages}
\usepackage{color}
\usepackage[pdftex,
	pdftitle={Titel},
	pdfsubject={},
	pdfauthor={Aothor1, Author2},
	pdfkeywords={},
	linkbordercolor=white,
        urlbordercolor=white,
        citebordercolor=white
]{hyperref}
\hypersetup{
urlcolor=black,
colorlinks=true,
linkcolor=black,
citecolor=black,
}%
\usepackage[utf8]{inputenc}
\usepackage{ifthen}
\usepackage{colortbl}
\usepackage{hhline}
\usepackage[absolute,showboxes]{textpos}
\usepackage[absolute]{textpos}
\usepackage{booktabs}
\usepackage{tabularx}
\usepackage{datatool}

\usepackage{tikz}
\usepackage{graphicx}
\usepackage{textcomp}
\begin{document}
\definecolor{tablered}{RGB}{255,199,206}
\definecolor{tableyellow}{RGB}{255,235,156}
\definecolor{tablegreen}{RGB}{198,239,206}
\newboolean{OPTIONgerman} 
 \setboolean{OPTIONgerman}{false} 
\ifOPTIONgerman
  \renewcommand{\contentsname}{Inhaltsverzeichnis}
  \renewcommand{\listfigurename}{Abbildungsverzeichnis}
  \renewcommand{\listtablename}{Tabellenverzeichnis}
  \renewcommand{\refname}{Literaturverzeichnis}
  \renewcommand{\indexname}{Stichwortverzeichnis}
  \renewcommand{\figurename}{Abbildung}
  \renewcommand{\tablename}{Tabelle}
  \renewcommand{\partname}{Teil}
  \renewcommand{\appendixname}{Anhang}
  \renewcommand{\abstractname}{Zusammenfassung}
  \renewcommand{\IEEEkeywordsname}{Schl\"usselw\"orter}
  \renewcommand{\IEEEproofname}{Beweis}
\else
  \renewcommand{\contentsname}{Contents}
  \renewcommand{\listfigurename}{List of Figures}
  \renewcommand{\listtablename}{List of Tables}
  \renewcommand{\refname}{References}
  \renewcommand{\indexname}{Index}
  \renewcommand{\figurename}{Figure}
  \renewcommand{\tablename}{Table}
  \renewcommand{\partname}{Part}
  \renewcommand{\appendixname}{Appendix}
  \renewcommand{\abstractname}{Abstract}
  \renewcommand{\IEEEkeywordsname}{Index Terms}
  \renewcommand{\IEEEproofname}{Proof}
\fi
\newcommand{\paper}{Analog Computing for the $21^\text{st}$ Century}
\title{Analog Computing for the $21^\text{st}$ Century}
\author{{Bernd Ulmann}}
\newcommand{\MPCdate}{Juni 2023}
\newcommand{\logo}{anabrid.png}
\newcommand{\ConfTitle}{ \parbox[b][14mm][t]{40mm}{\raggedright{\MakeUppercase{MPC-Workshop \MPCdate}}}} 
\newcommand{\DocTitle}{  \parbox[b][14mm][t]{60mm}{\raggedleft{\MakeUppercase{\paper} }}}
\newcommand{\logoright}{\ConfTitle\hfill\includegraphics[height=14mm, keepaspectratio]{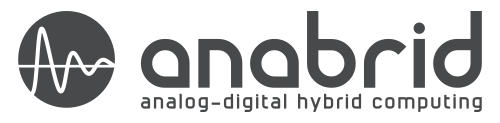}}
\newcommand{\logoleft}{\includegraphics[height=14mm, keepaspectratio]{\logo}\hfill\DocTitle}
\markboth{\logoleft}%
{\logoright}
\maketitle
\begin{abstract}
Many people think of analog computing as a historic dead-end in computing.
In fact, nothing could be further from the truth as analog computing -- together 
with quantum computing -- has the potential to bring computing to new levels with
respect to raw computational power and energy efficiency. The following
paper explains the limits of digital computers, gives a quick introduction to 
analog computing in general, and shows a number of recent developments that
will change the way we think about computers in the next few years.
\end{abstract}
\section{Introduction}
 After many decades of extremely successful application of stored-program
 digital computers (just called \emph{digital computers} here)
 two alternative computing paradigms are shifting into the limelight: Analog 
 computing and quantum computing. Both hold much promise for the future.
 The reasons for this forthcoming paradigm shift away from
 digital computers, the basic ideas of analog computing, application areas,
 etc. are described here.

 It should be noted here that \emph{analog computing}, despite its long 
 history (much longer than that of digital computers) is a very active and
 promising area of research. One should not think of large classic analog
 computers in a museum but instead of modern integrated circuits and high 
 performance computing. The $21^\text{st}$ century needs analog computers
 on integrated circuits as co-processors employing the current power 
 efficient CMOS technology, so we are talking about cutting-edge technology,
 not some arkane relics of the past.
\section{Why we should care about analog computing}
 Despite their incredible success and versatility, digital computers are about
 to hit fundamental physical limits and face a variety of other challenges thus
 shifting the focus of current research to different computational
 paradigms.

 The first problem to mention is the high power consumption of our digital
 infrastructure, a large part of which is caused by the very digital computers
 at its heart. The overall ICT (\emph{Information and Computing Technology})
 sector is estimated to have consumed between 4\% and 6\% of the global electrical
 energy in 2020 with data centers accounting for about one sixth of that total 
 number.\footnote{See \cite{POSTNOTE}.} In particular the demand for CPU power 
 for training deep artificial neural networks is growing at an ever increasing 
 pace, thus further advancing the energy demands of our current computing 
 infrastructure.  Apart from the obvious implications for future systems and data 
 centers with respect to energy consumption, this creates an additional problem --
 that of heat removal. Modern top performing CPUs such as members of AMD's 
 \emph{Ryzen Threadripper} family have TDP (\emph{Thermal Design Power}) values 
 of up to $250$ Watt. Cooling large scale systems based on such chips is no 
 easy task.

 Interestingly, clock frequencies of digital computers have not increased much 
 at all during the last two decades which is largely due to the fact that 
 the overall power consumption of digital circuits increases super-linearly with
 clock frequency. Admittedly the energy efficiency of these systems has risen 
 several times due to advances in computer architecture and a trend towards 
 many-thread-architectures coupled with more and more specialised subsystems but
 these speedups are neither easy to implement nor applicable to all 
 applications.

 Generally, parallelism is rather hard to achieve in digital computers as 
 observed and described by \textsc{Gene Myron Amdahl} as early as 1967. Even if 
 \emph{\textsc{Amdahl}'s law}, derived in this seminal publication, is an 
 oversimplification and does not take the structure of modern processors with 
 their intricate caching systems, etc., into account, it turns out to be a rather 
 good tool for estimating the maximum speedup achievable by parallelising a 
 task.\footnote{See \cite{AMDAHL}.} Admittedly, there is a variety of tasks which 
 benefit rather directly from parallel processing as described by 
 \emph{\textsc{Gustafson}'s law}\footnote{See \cite{GUSTAFSON}.} but this is
 not a general rule. To get near the theoretical peak-performance of a given
 system with a real-world application is not a simple task, often leaving 
 large parts of a CPU sitting idle. 

 Current technology nodes of $2$ nm are about to hit fundamental boundaries 
 of integration densities, and of the many billions of transistor functions in 
 a modern CPU, typically only a small part is actually performing computations 
 while the vast majority implements infrastructure such as high-speed caches, 
 intricate control circuits ranging from out-of-order-execution to complex 
 pipeline control,
 register renaming, a plethora of uncore functions, bus and memory interfaces,
 and many more. At the very heart of these systems are comparatively few ALUs 
 (\emph{arithmetic logic unit}s) performing actual computations, which does not 
 make really good use of the available silicon real estate.
 
 A comparison of two state-of-the-art supercomputers might be interesting:
 Today's fastest digital supercomputer, \emph{Frontier}, a HPE Cray EX system
 located at the Oak Ridge National Laboratory, achieves a staggering 
 $1,194$ PFLOPS ($=1.194\cdot10^{18}$ floating point operations per second, 
 measured with the LINPACK benchmark) using $8,699,904$ cores and consuming
 about $22.7$ MW of electrical power. This amounts to 52 GFLOPS per Watt which
 is really impressive but next to nothing compared with the contender, a human
 brain. The raw computational power of a brain is estimated to be in the ball
 park of $38$ PFLOPS at an overall power consumption of about 20 Watt. This
 amounts to about $1.9$ PFLOPS/W -- many decades better than our current
 digital supercomputers.

 How is that possible? What does nature do so differently from our classic
 approach to computation to achieve such an extraordinary high energy 
 efficiency? It is mainly the fact that the brain does not execute an 
 algorithm, i.\,e., there is no sequence of instructions to be executed in a
 step-by-step fashion. Instead a brain consist (vastly simplified) of billions
 of rather simple computing elements, neurons, which are interconnected in a 
 suitable fashion to implement the many feats we are capable of. Thus, the
 \emph{program} underlying a brain is not an algorithm but a directed graph 
 describing the interconnection of neurons with their individual input 
 weights etc. There is no central memory, no cache memory, no intricate 
 control system, just a multitude of small computing elements all working
 in full parallelism.

 From that perspective, a biological brain quite closely resembles an 
 \emph{analog computer} instead of a stored-program digital computer, as the
 interconnection of its computing elements is the actual program.
\section{Analog computing}
 Now, what is an analog computer? In a nutshell, an analog computer consists of 
 a number of computing elements, each capable of performing a basic 
 mathematical operation such as summation, integration (this is a truly magic
 element), multiplication, etc., which are interconnected in a suitable way
 to form a model, an \emph{analogue}, for a given problem to be solved. The 
 term \emph{analog computer} refers to the model building charactersitic and not
 to a certain representation of values.

 Variables are typically represented by voltages or currents.\footnote{It should
 be noted that there are even ``digital analog computers'' called \emph{DDAs} 
 (\emph{Digital Differential Analyser}s) which represent values in a binary 
 fashion but also rely on a set of computing elements which are interconnected
 with each other to implement a program.} This representation simplifies the 
 construction of an analog computer substantially since only a single wire
 (or a pair of wires) is required to connect two computing elements. Nothing is
 more true in computer science than the saying ``there is no such thing like a 
 free lunch'' which also holds true in the case of analog computers where this 
 characteristic comes at a cost: An analog computer typically offers only limited 
 precision with typical resolutions being of the order of $10^{-3}$ to $10^{-4}$.
 Analog computers capable of a precision of $10^{-4}$ are often called 
 \emph{precision analog computers}. 

 As negative as this sounds it isn't much of a problem for many if not most 
 applications. Engineering problems typically do not require more than a few 
 decimal places for a useful solution, biological neural networks only feature
 very limited resolution for their synaptic weights which thus also holds true
 for artificial neural networks, etc. In cases where higher precision is required,
 results generated by an analog computer could be used as starting points for 
 numerical algorithms executed on an attached  digital computer which can then 
 enhance these results to the required degree of precision.

 Using voltages or currents to represent values not only removes much 
 complexity from a computer but also adds significantly to the energy 
 efficiency of the system since there is no need to flip zillions of signal lines 
 between $0$ and $1$ billions of times per second, each time charging or 
 discharging tiny parasitic capacitors. Furthermore this representation 
 facilitates interfacing to the surrounding world making analog computers ideal 
 for signal pre- and postprocessing, etc.

 Analog computers are ideally suited for problems that can be described as 
 (systems of coupled) differential equations (most problems relevant in 
 science and engineering fall into this category). There are also suitable 
 approaches to tackle partial differential equations on analog computers. They 
 can even implement oscillator based \emph{\text{Ising} machines} and thus solve 
 problems which are normally attributed to adiabatic quantum 
 computers.\footnote{See \cite{CHOU}, \cite{BASHAR} etc.}

 The following simple problem shows the main difference between programming
 a classic digital computer and an analog computer. Here, $x=a(b+c)$ is to be
 computed. A classic digital computer requires six instructions to accomplish
 this as shown in figure \ref{pic_xabc_digital}. The two arithmetic operations
 at the core of the task are surrounded by four instructions to load data from
 memory and store the result back to memory. Of course, this ratio gets better
 with increasing problem complexity and with clever register allocation schemes, 
 but it illustrates the nature of a digital computer quite well.
 \begin{figure}
  \centering
  \begin{verbatim}
LOAD  A, R0
LOAD  B, R1
LOAD  C, R2
ADD   R1, R2, R1
MULT  R0, R1, R0
STORE R0, ...
  \end{verbatim}
  \vspace*{-5mm}
  \caption{Computing $x=a(b+c)$ on a digital computer}
  \label{pic_xabc_digital}
 \end{figure}

 Solving the same problem on an analog computer requires two computing elements,
 one summer and one multiplier, and a few connection between these elements. The 
 input values $a$, $b$, and $c$ represented by voltages or currents are connected 
 to these elements, while the output of the summer is connected to one input of 
 the multiplier. There is no algorithm in the classic sense of the word, instead 
 the program resembles a directed graph describing how computing elements must be 
 connected with each other to solve a given problem. Since there is no algorithm 
 and no central memory storing instructions and values, all computing elements 
 can work in full parallelism with no need for explicit synchronisation etc.
 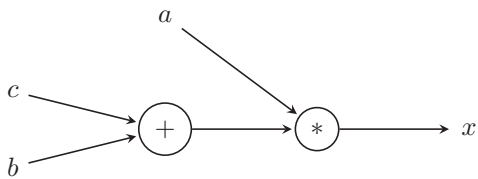
\begin{figure}
  \centering
  \begin{tikzpicture}[
    > = stealth, 
    shorten > = 1pt, 
    auto,
    node distance = 3cm, 
    semithick 
   ]

   \tikzstyle{every state}=[
    draw = black,
    thick,
    fill = white,
    minimum size = 4mm
   ]

   \node (B) at (0, 0) {$b$};
   \node (C) at (0, 1) {$c$};
   \node (PLUS) at (2, 0.5) [circle, draw] {$+$};
   \draw[->] (B) to (PLUS);
   \draw[->] (C) to (PLUS);
   \node (MULT) at (4, .5) [circle, draw] {$*$};
   \draw[->] (PLUS) to (MULT);
   \node (A) at (2, 2) {$a$};
   \draw [->] (A) to (MULT);
   \node (RESULT) at (6, .5) {$x$};
   \draw[->] (MULT) to (RESULT);
  \end{tikzpicture}
  \caption{Analog computer setup for solving $x=a(b+c)$}
  \label{pic_xabc}
 \end{figure}

 This leads to an interesting basic difference between digital and analog 
 computers: In a digital computer it is always possible to trade problem 
 complexity against solution time. The more complex a problem gets the
 longer it will take to solve it on a digital computer (always assuming there is
 enough memory to hold all instructions and variables for the problem). This
 often convenient tradeoff is not possible with an analog computer, which is a 
 certain drawback. If the implementation of a problem requires more computing
 elements than a given analog computer contains, it cannot be solved (at least 
 not directly) on this particular machine. On the other hand, the solution time 
 for a problem on an analog computer does not increase with problem size,
 provided that there are enough computing elements available to implement the 
 program.

 Of course, there are some drawbacks of analog computers: In addition to the 
 rather limited precision of analog computing elements, values have to be within 
 the interval $[-1,1]$.\footnote{It is best to always think in terms of this 
 abstract interval instead of the actual minimum and maximum voltages or currents 
 representing these values in order to simplify programming and scaling. 
 Early vacuum tube based analog computers used voltages between $\pm100$~V to
 represent values. These voltages later dropped to $\pm10$~V in transistorised
 machines, while modern implementations use $\pm1$~V and even lower voltages at 
 much increased bandwidths.} Additionally, the generation of arbitrary functions 
 as obtained by experiments or the like is also quite challenging for an analog 
 computer especially when functions $f(x_1,\dots,x_n)$ of more then one argument 
 are required. Nowadays this can be easily overcome by using lookup tables stored
 in some memory with attached analog-digital- and digital-analog-converters 
 (\emph{ADC}s and \emph{DAC}s). Regarding integration, being a basic operation 
 of an analog computer, only time is available as the free variable of 
 integration; this constraint requires the use of advanced techniques to 
 tackle PDEs (\emph{partial differential equations}).

 Modern analog computers typically will be part of a \emph{hybrid computer}, 
 i.\,e., a combination of a digital computer with an analog computer as a 
 specialised, closely attached co-processor. This co-processor excels at the 
 high-speed, energy efficient solution of (systems of) differential equations 
 while the digital computer allows for storage, decision making, function 
 generation, parametrisation of the analog computer, etc.
\section{Programming}
 As arcane as analog computer programming may look to the uninitiated, it is 
 much more straight forward than the algorithmic approach we were all taught
 and cherish. A simple example may show the basic approach which
 relies on an idea of Lord \textsc{Kelvin} who developed the 
 \emph{\textsc{Kelvin} feedback technique}\footnote{See \cite{KELVIN_FEEDBACK}.}
 in 1876 after the invention of a practical mechanical integrator by his 
 brother.\footnote{See \cite[pp.~22~ff.]{ULMANN_AC}.}

 This technique transforms a mathematical problem description into 
 an analog computer program, i.\,e., a directed graph connecting suitable 
 computing elements, in a series of five steps:
 \begin{enumerate}
  \item Organise the equation to isolate the highest derivative on the left
   hand side.
  \item Assuming this derivative is known, all lower derivatives can be 
   generated by repeated integrations.
  \item Using summers, multipliers, and other computing elements, all terms
   on the right hand side of the equation obtained in the first step are 
   derived.
  \item These terms are now tied together to form the right hand side of the 
   equation. Since this must be equal to the highest derivative this signal is 
   then fed back into the circuit as the highest derivative which was assumed
   to be known in the second step.
  \item This program typically needs scaling to ensure that no value lies
   outside the interval $[-1,1]$. In addition to this, a good scaling process
   also ensures that ideally every variable will make best use of this 
   interval, thus increasing the precision of the computation.
 \end{enumerate}

\begin{figure}
  \centering
  \includegraphics[width=\columnwidth]{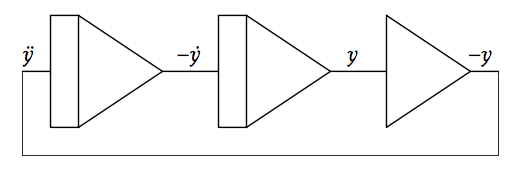}
  \caption{Basic analog computer setup for solving $\ddot{y}+y=0$}
  \label{pic_sine_1}
 \end{figure}

 Suppose the $2^\text{nd}$-order differential equation $\ddot{y}+y=0$ is to 
 be solved with an analog computer. It first is solved for its highest
 derivative yielding $\ddot{y}=-y$. Starting from $\ddot{y}$ the lower 
 derivatives $-\dot{y}$ and $y$ are derived by a chain of two 
 integrators. (Due to the actual implementation of such integrators,
 they typically perform an implicit sign flip. Summers behave similarly.)
 Since the right-hand side of $\ddot{y}=-y$ is negative, an additional 
 change of sign is required which is done by a summer with only one input,
 acting as an inverter. The output of this summer is then fed into the 
 first integrator, thus completing a feedback loop as shown in figure 
 \ref{pic_sine_1}. The triangular symbol on the right denotes a summer while
 the two symbols to its left represent integrators.

\begin{figure}
  \centering
  \includegraphics[width=\columnwidth]{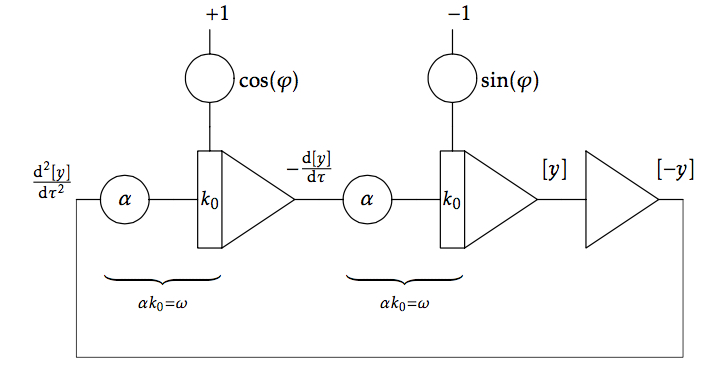}
  \caption{Analog computer setup for solving $\ddot{y}+\omega^2y=0$ with 
   proper initial conditions}
  \label{pic_sine_2}
 \end{figure}

 This program is missing one important detail -- the initial conditions
 of the integrators, specifying which solution to obtain. A complete 
 program for the solution of the equation $\ddot{y}+\omega^2y=0$ with initial
 conditions $y(0)=\sin(\varphi)$ and $\dot{y}(0)=\cos(\varphi)$ is shown in 
 figure \ref{pic_sine_2}. The circles denote coefficients, $k_0$ denotes the 
 \emph{time-scale factor}\footnote{This time-scale factor $k_0$ describes the 
 speed at which an integrator runs. In the case of $k_0=1$ the output of an 
 integrator will reach $-1$ with a constant input of $+1$ after one second. If
 $k_0=10^3$ this value will be reached after one millisecond, etc.} of the 
 integrator with $\alpha k_0=\omega$.

 An analog computer typically features (at least) three modes of operation: 
 \begin{itemize}
  \item \emph{Initial Condition} mode (\emph{IC}): The integrators are set to 
   their respective initial condition values. This step must precede the following
   mode.
  \item \emph{Operate} (\emph{OP}): The integrators integrate with respect to 
   time -- the computer performs its actual computations.
  \item \emph{Halt} mode (\emph{HALT}): All integrators are halted, so that all 
   variables within the program remain constant. This is typically used to read 
   values when using slow ADCs. This mode may be followed by OP or IC.
 \end{itemize}

 Problems of a more realistic complexity can be transformed into analog computer
 programs employing the same rationale but typically need some (extensive) 
 attention with respect to scaling.\footnote{A recent and thorough introduction 
 to analog and hybrid computer programming can be found in \cite{ULMANN_AHCP}.}
\section{Problems of classic analog computers}
 While analog computers were the systems of choice when it came to the 
 treatment of problems requiring real-time solutions of highly complex 
 dynamic systems, they were basically replaced by digital computer in the 
 early 1980s even though their digital rivals could not compete 
 their computational power for another one or two decades to come. 

 The main reasons for this were that digital computer became quickly cheaper while
 analog computers continued to be rather expensive systems due to the 
 required high precision components, etc. In addition to this, programming a
 classic analog computer required manually plugging hundreds and sometimes
 thousands of patch cables on a patch panel to implement a certain analog 
 computer program, a time consuming and error prone process. Although such
 patch panels with their intricate maze of wires could be quickly replaced on 
 medium to large scale analog computers, switching from one program to another
 still was a time consuming affair. Accordingly, most analog computers were used 
 to treat a single problem for a prolonged time during which no other program 
 could be run concurrently. In contrast to this digital computers offered time 
 sharing access since the 1960s allowing many users to share the (limited) 
 computing power of these machines. This characteristic alone often turned 
 purchase decisions away from an analog computer. 

\begin{figure}
  \centering
  \includegraphics[width=\columnwidth]{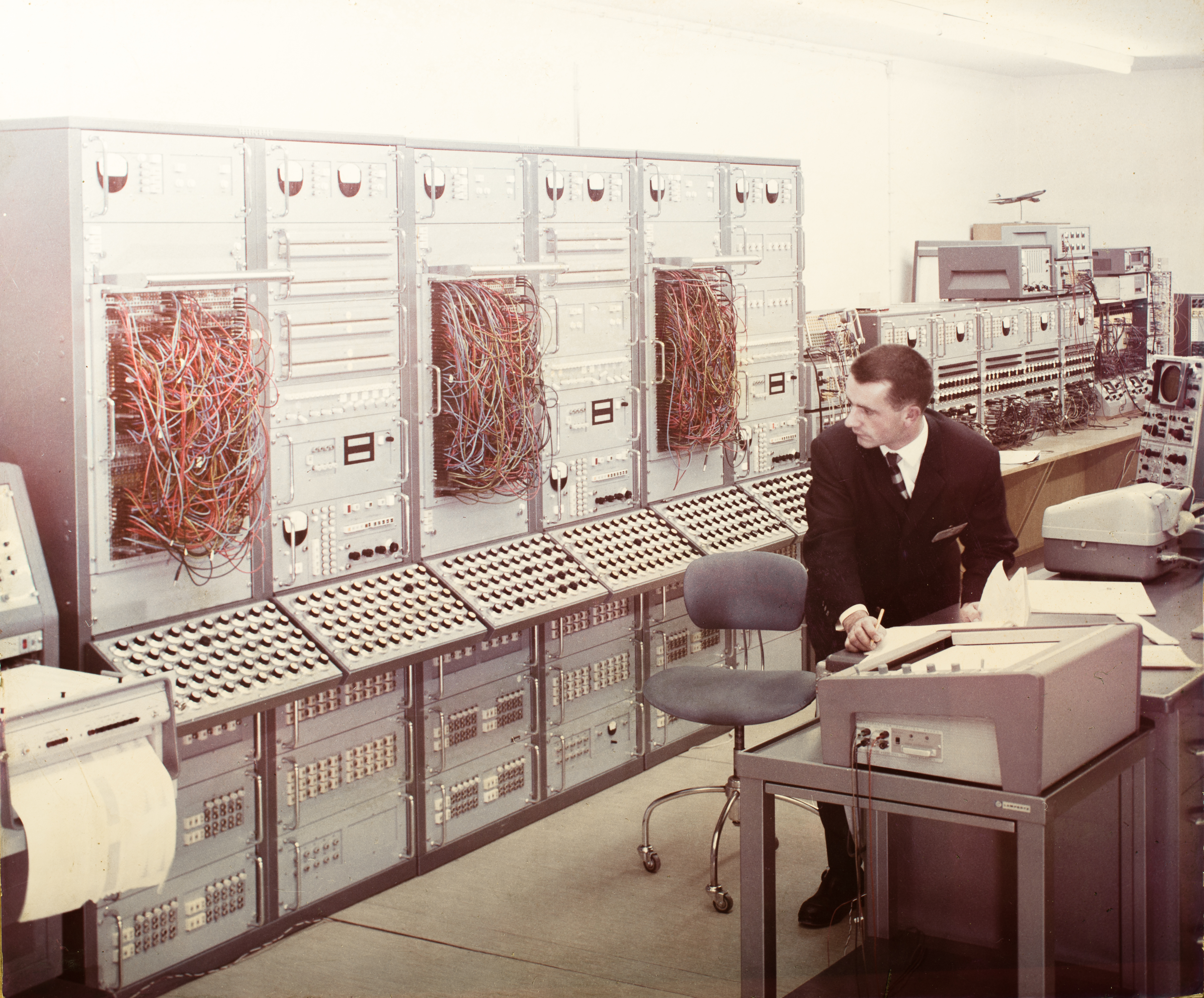}
  \caption{Classic large scale analog computer installation}
  \label{pic_boelkow}
 \end{figure}

 Figure \ref{pic_boelkow} shows a classic large scale analog computer 
 installation at the German aerospace company Boelkow in 1960. This particular
 installation implemented a flight simulator for a vertical-takeoff-and-landing
 jet fighter. It shows clearly the fact that the analog computer's size must
 match the problem size. At its heart are three large Telefunken RA800 computers 
 as well as a number of smaller scale table top analog computers, all being 
 interconnected to form a single very large scale machine. It is quite impressive 
 that it was possible to implement a realistic flight simulator controlling
 a hydraulic hexapod with a cockpit mounted on its top in 1960 using this 
 approach, something digital computers needed many years to actually compete 
 with. 

 At the same time this picture clearly shows how cumbersome programming these 
 systems was back then. The patch panels are buried under heaps of patch cables 
 connecting hundreds of computing elements \mbox{with} each other. Changes to 
 such a 
 complicated setup often required hours or even days and sometimes weeks of 
 preparation and actual implementation. Changing parameters of a program required 
 manually changing hundreds of precision 10-turn potentiometers, $300$ of which 
 are visible on the three large systems on the left in the figure with many more 
 on the smaller machines on the right.\footnote{A detailed history of analog 
 computers can be found in \cite{ULMANN_AC}.}
\section{Areas of application}
 There is a wide variety of applications for modern analog computers. The most 
 prominent of which are the implementation of artificial neural networks by 
 means of analog electronic implementations of neurons, and high performance 
 computing 
 with applications such as molecular dynamics, computational fluid dynamics, 
 Monte-Carlo simulations as often used in financial mathematics, optimisation 
 tasks, and many more. These fields can benefit most from the high computational
 power of analog computers with their high energy efficiency in second place.

 Another large application area where the high energy efficiency is of utmost
 importance are medical applications such as cardiac and brain pacemakers, 
 blood sugar sensing and insulin pump control, in vivo instrumentation for 
 cancer therapy, and many more. In some specialised applications it might be 
 possible to power the implant by energy harvesting thus alleviating the need
 for bulky and inconvenient energy storage solutions, much to the benefit of 
 the patient.

 Since an analog computer does not execute an algorithm and has no traditional
 memory containing instructions or the like it is not prone to classic attack
 vectors, making it a worthwhile option for process control in critical 
 environments and the like.

 Used for signal processing, analog computers can perform trigger word 
 detection for smart devices, vibration analysis, and other tasks for 
 predictive and preventive maintenance, etc.
\section{Modern technology}
 Today, the main drawbacks of these classic systems can be overcome using 
 modern CMOS technology. The patch panel is a thing of the past and belongs
 in a museum. Modern analog computers will be fully reconfigurable using
 large switch matrices (\emph{crossbar switches}) under the control of an
 attached digital computer. Also the manual potentiometers have long since
 been replaced by digital potentiometers or multiplying DACs allowing the 
 digital computer to change parameters in microseconds.

 In addition to this, modern implementations of analog computers will offer 
 much higher bandwidth computing elements, thereby allowing for much shorter 
 solution 
 times. Combined with automatic reconfiguration and reparametrisation this will 
 make it possible to use an analog computer in a time-sharing like mode of 
 operation.
 
 One central problem that must be addressed in order to get widespread 
 acceptance 
 for analog and hybrid computing is the necessity of an abstract programming 
 language, a \emph{domain specific language} (\emph{DSL}), that must hide most 
 if not all of the underlying physical principles the analog computer
 relies on. The reason for this is the fact that programming analog computers
 requires a completely different mindset to writing an algorithm. Since 
 the majority of programmers and IT professionals are trained to work in the
 algorithmic domain, the impedance mismatch between this approach and that 
 of specifying the interconnection of individual computing elements must be
 minimised.

 Furthermore, it is not that simple to tightly couple an analog computer with
 a digital system, especially since the analog computer offers solution times
 in the sub-microsecond range. This requires very short interrupt latencies on 
 the digital computer so that it can keep up with its analog co-processor without 
 forcing it to idle most of the time waiting for some interrupt to be processed.
\section{Modern developments}
 This section gives a quick overview over recent developments in the field of
 analog and hybrid computing including academic and commercial projects.

 Terminating a long hiatus in the field of analog computing, \textsc{Glenn 
 E. R. Cowan} developed and implemented a reconfigurable analog computer on a 
 VLSI chip in 2005.\footnote{See \cite{COWAN}.} Its development showed the 
 basic feasibility of CMOS based analog computers. This was then followed by an 
 enhanced VLSI analog computer chip developed and implemented by \textsc{Ning
 Gou} in 2016.\footnote{See \cite{GUO}.} Unfortunately neither of these 
 academic projects resulted in commercial developments. The bandwidth of the 
 computing elements was rather limited with sometimes large phase shifts caused 
 by the interconnect matrices. Also the software support was quite limited.

 An early commercial development was due to \emph{anadigm}\textregistered\ who 
 developed and market \emph{FPAA}s (\emph{Field Programmable Analog Arrays})
 consisting of a number of \emph{Configurable Analog Block}s (\emph{CAB}s) 
 and a number of input/output interfaces. These can be used to implement a
 variety of signal pre- and postprocessing tasks such as filters, general audio 
 signal conditioning, etc.\footnote{A nice signal processing board based on 
 such an FPAA was developed and is sold by \textsc{Nicolas Steven Miller}
 (\url{https://zrna.org}). These boards can be easily configured using a 
 Python client.}

 \emph{Aspinity} has developed two major technologies: The 
 \emph{RAMP\texttrademark\footnote{\emph{Reconfigurable Analog Modular 
 Processor}} Technology Platform}, a programmable analog neuromorphic processor,
 and the \emph{AnalogML\texttrademark Core}, a machine learning co-processor.
 Typical applications are trigger word detection for smart devices, glass 
 break detection, acoustic event detection, and vibration monitoring. Due
 to the high energy-efficiency these devices can be applied in always-on 
 sensing applications. 

 Another startup working towards analog computing for artificial intelligence is 
 \emph{Mythic} (\url{https://mythic.ai/}). They pioneer a \emph{compute-in-memory}
 approach where memory cells are implemented as variable resistors in the form of 
 flash memory cells. A matrix of such cells can perform typical operations of 
 linear algebra by employing \textsc{Kirchhoff}'s law, being fed row wise by DACs 
 with ADCs reading out the results of the implicit additions and multiplications
 performed by the resistor array. 

 \emph{IBM}, too, has developed an analog AI accelerator, which was unveiled in 
 2022.\footnote{\url{https://research.ibm.com/blog/the-hardware-behind-analog-ai}}
 It uses \emph{Resistive RAM} (\emph{ReRAM}) made from \emph{Phase Change 
 Memory} (\emph{PCM}) cells the state of which is switched between an 
 amorphous and a crystalline state. This experimental chip implements 
 $35\cdot10^6$ such PCM cells. 

 A very interesting development is the implementation of artificial neural 
 networks in a true three-dimensional topology. Stacking of chips has been 
 done for many years in the digital domain but is inherently limited in its
 extent due to the high power consumption and the problems of removing the 
 excess heat from such a stack of chips. The very high energy-efficiency of
 analog computing approaches will make large scale three-dimensional 
 topologies feasible.\footnote{See \url{https://research.ibm.com/blog/vlsi-hardware-roadmap}.}

 One might ask about the application of Memristors as they are often mentioned
 being ``ideal'' devices for implementing synaptic weights in analog neural
 networks, etc.\footnote{ReRAM cells are typically not considered being 
 Memristors due to their rather different behavior.} It may be doubted that 
 Memristors based on filament conduction would be a good choice for implementing
 synaptic weights due to their typically limited number of state changes as well
 as the necessity for a device forming period prior to their actual use in a circuit. Approaches to
 Memristors not relying on filament conduction are relatively new and seem
 to have not yet matured into integration-ready devices on a large scale.
 However, using Memristors in analog artificial neural networks might be 
 interesting as they might be able to implement self-learning systems based 
 on assumptions such as ``fire together, wire together''.

 Apart from these rather specialised and mostly AI centric analog computing
 applications and approaches, there is another startup, \emph{anabrid} GmbH, based
 in Germany\footnote{The author is one of the founders of anabrid GmbH.} 
 pursuing general purpose analog computing with the ultimate goal of designing,
 implementing, and marketing a reconfigurable large-scale integrated analog 
 processor. This device will act as a co-processor offloading compute intensive
 tasks from a classic digital processor but without being restricted to a 
 narrow field of application, which is in contrast to the aforementioned
 developments.

 Analog computing like quantum computing comes with the challenge that 
 programming such machines is completely different to the classic algorithmic
 approach taught in schools and universities. Especially professional 
 programmers have lots of difficulties addapting to a different computational
 paradigm like that of an analog computer. To bridge this gap at least two
 things are required: First, a cheap analog computer aimed at the educational
 market, at hobbyists, etc. Second, an abstract domain specific programming 
 language which allows the seemless integration of analog co-processors  into 
 current digital computer systems. Ideally, a programmer should not have to
 think about the actual implementation of an analog computer or of intricacies
 such as scaling and the like. 

\begin{figure}
  \centering
  \includegraphics[width=\columnwidth]{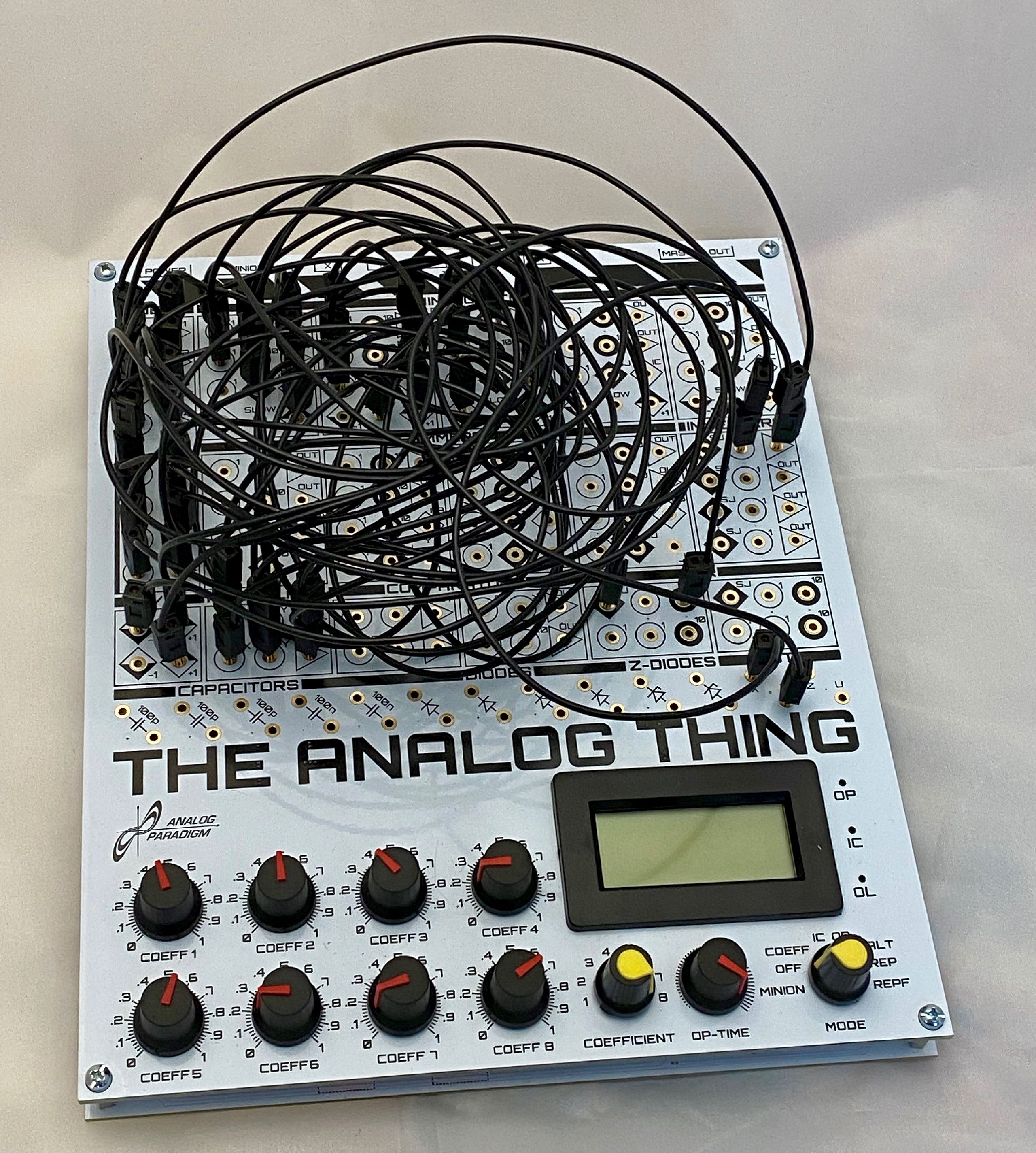}
  \caption{THE ANALOG THING}
  \label{pic_that}
 \end{figure}

 Figure \ref{pic_that} shows \emph{THE ANALOG THING}, an open hardware 
 project\footnote{See \url{https://the-analog-thing.org} and 
 \url{https://github.com/anabrid/the-analog-thing} for schematics etc.} aimed at
 the educational market. This little analog computer contains enough computing 
 elements for serious experiments in analog computing and can be easily interfaced
 to a digital computer, thus forming a hybrid computer.\footnote{See 
 \url{https://the-analog-thing.org/wiki/Hybrid_Computer}} It contains five
 integrators, four summers, four sign-inverters, two multipliers, two 
 comparators, and eight manual coefficient potentiometers. Of this system
 more than 1000 have already been ordered showing a strong interest in 
 analog computing in and for the $21^\textsc{st}$ century.
\section{Conclusion}
 It may be concluded that analog computing is making a comeback and will stay
 with us in the future. Analog computers, ranging from specialised devices to 
 general purpose co-processors will substantially transform the way we compute
 in general. The following years will see huge leaps in analog computer 
 implementations catching up with the developments in the digital domain during
 the last decades, eventually surpassing our classic computers for certain 
 areas of application.

%

\renewcommand{\BiographyAddLine}{1}
\begin{IEEEbiography}[{\includegraphics[width=1in, height=1.25in, clip,keepaspectratio]{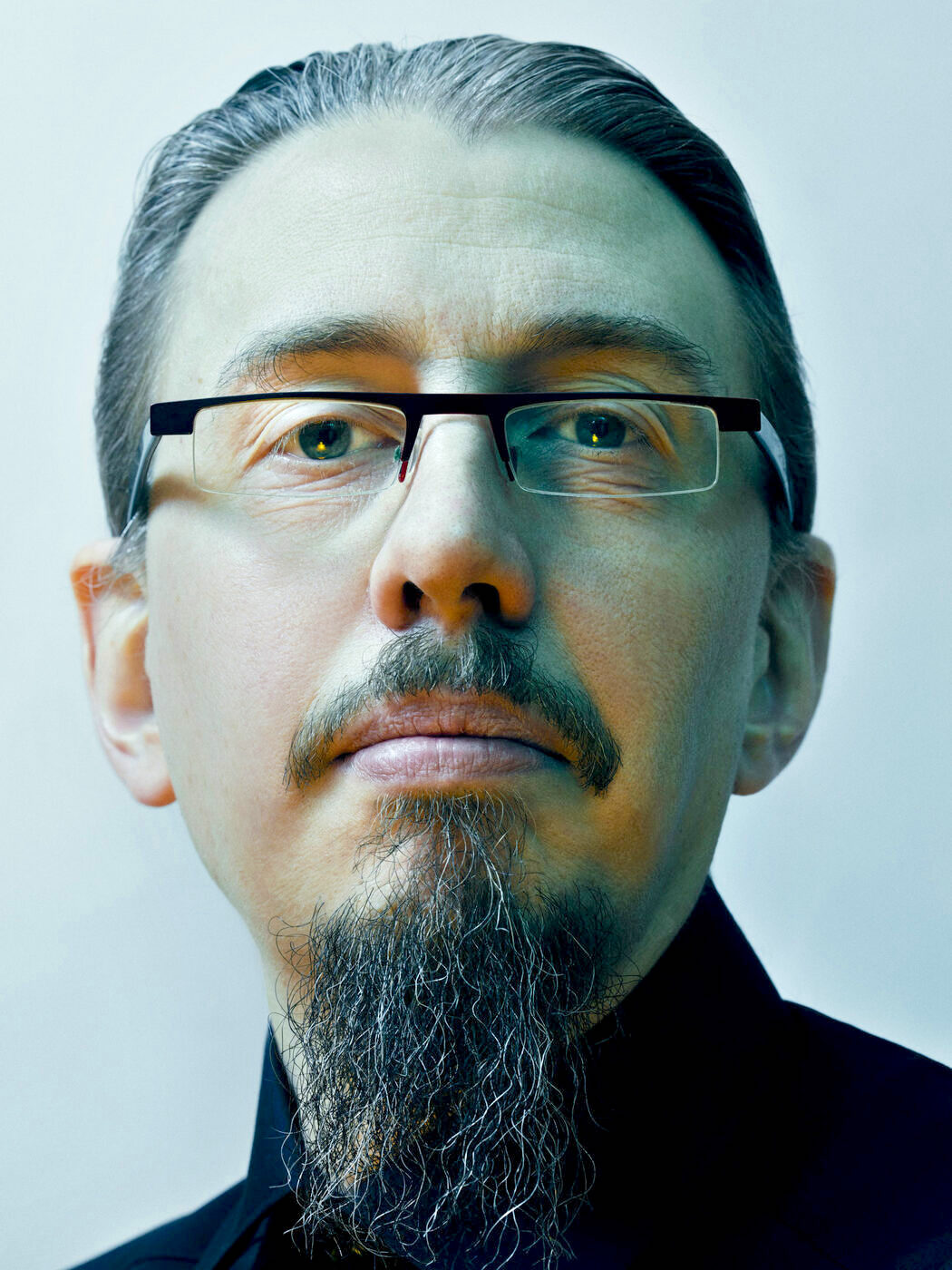}}]
{Bernd Ulmann}
Bernd Ulmann is co-founder of anabrid GmbH and professor for business computer 
science at FOM University of Applied Sciences for Economics and Management. He
studied mathematics with a minor in philosophy at the Johannes Gutenberg 
university Mainz. His main interests are analog and hybrid computing and the 
simulation of dynamic systems. He also collects and restores classic analog 
computers and has a soft spot for chaotic systems and their implementation on 
analog computers.
\end{IEEEbiography}
\renewcommand{\BiographyAddLine}{1}
%
%
%
\end{document}